\documentclass{article}
\usepackage{spconf,amsmath,graphicx}
\usepackage{cite}
\usepackage{makecell}
\usepackage{amsfonts}
\usepackage{multirow}
\usepackage{hyperref}
\usepackage{fancyhdr}   
\usepackage{graphicx} 
\graphicspath{ {images/} }

\title{SPECTROGRAMS ARE SEQUENCES OF PATCHES }
\name{Leyi Zhao, Yi Li\thanks{A great thanks to Li Shiqian at Virginia Tech for his assistance in the experiment implement}  } 

\address{Sichuan University}

\begin{document}
%
\maketitle
\begin{abstract}
Self-supervised pre-training models have been used successfully in several machine learning domains. However, only a tiny amount of work is related to music. In our work, we treat a spectrogram of music as a series of patches and design a self-supervised model that captures the features of these sequential patches: Patchifier, which makes good use of self-supervised learning methods from both NLP and CV domains. We do not use labeled data for the pre-training process, only a subset of the MTAT dataset containing 16k music clips. After pre-training, we apply the model to several downstream tasks. Our model achieves a considerably acceptable result compared to other audio representation models. Meanwhile, our work demonstrates that it makes sense to consider audio as a series of patch segments.
\end{abstract}
\section{Introduction}
\label{sec:intro}

Self-supervised learning has been studied in many directions, including NLP, CV, and data mining. With the powerful transformer performance, pre-training models by self-supervised learning and then fine-tuning downstream tasks have become an NLP training paradigm. After the emergence of the Vision Transformer (ViT) \cite{dosovitskiy2020image}, more and more pre-trained models in the CV domain are also using transformers, so one can make these models work well in a specific task with a small amount of labeled data. In the music domain, labeled data is also challenging to obtain, so a model that can be pre-trained in unlabeled data by self-supervised learning is also of interest.

In numerous audio-related works, researchers have processed audio from a holistic direction. Among them, \cite{spijkervet2021contrastive,wu2021multi,castellon2021codified} are trained by directly treating the audio wave as pure sequence information, while \cite{zhao2022s3t,manco2022learning,Lin2022ContrastiveFL} are trained by converting the audio to a spectral signal and treating it as 2D pure image data. They all achieved excellent results. However, few works combine audio work with both NLP and CV, so we focus on proposing an approach that takes the best of both areas and can be pre-trained on large amounts of cheap unlabeled data and used for other downstream tasks

In computer vision, ViT partitions a picture into different patches, similarly processes them as tokens in NLP, and sends these patches inside a transformer for learning representations. In the music domain, a spectrogram can be regarded as a long strip of image data, and then we can divide it in a form as it is shown in fig1. As a result, the model learns not only the local intrinsic features of the spectrogram but also the relationships between the whole.

In our work, a two-stage pre-trained model is proposed, which contains a CNN-based feature extractor, a transformer-based bottleneck, and a CNN-based decoder. In the first stage In the first stage, the feature extractor and decoder are first trained by sampling different patches, ignoring the bottleneck, and training via autoencoding in order to let them learn the local features of the spectrogram. In the second stage, the entire spectrogram is sliced into segments of equal shape and then fed into the encoder trained in the first stage to obtain the features, which are learned in a Bert-like way, and later fed into the decoder for reconstruction.

Our work successfully combines NLP and CV algorithms by proposing a novel pre-trained model for music representation. We demonstrate that our model performs well in several downstream tasks with a minimal pre-training dataset and a guaranteed lightweight model. We also aim to provide support for other, more diverse MIR downstream tasks. Code is available at \href{https://github.com/annihi1ation/Patchifier-Neo}{https://github.com/annihi1ation/Patchifier-Neo} 
\begin{figure*}
	\centering
    \includegraphics[width=1\textwidth]{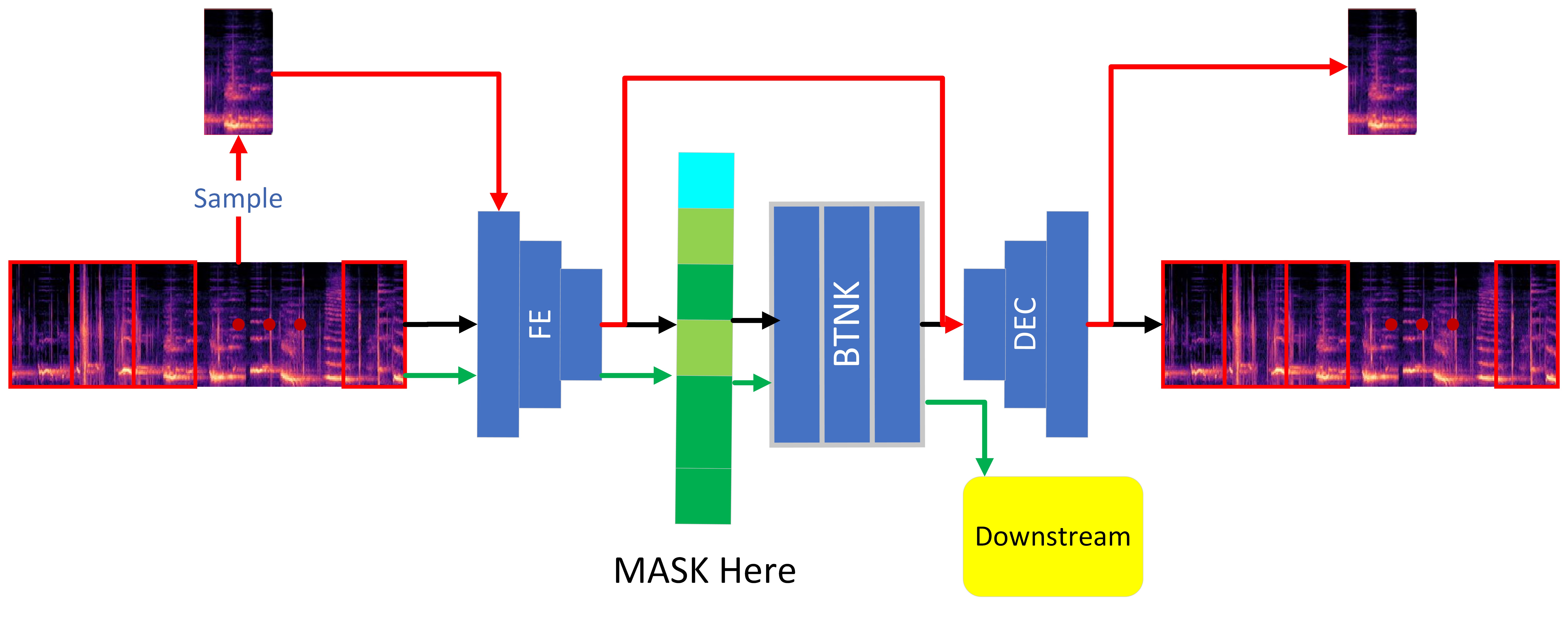}
    \caption{This is the patchifier's pipeline, where the red line represents the data flow for stage I. The black line is for stage II. The green line is for downstream task representation, azure squares are CLS tokens, light green is the masked tokens, and dark green is the embedding learned by the feature extractor. }
    \label{fig:backbone}
\end{figure*}
\section{Related Work}
\label{sec:format}

With the development of deep learning, models are becoming more and more capable of representation. However, labeled data is still challenging to obtain. Therefore, it is REASONABLE to learn valuable representations with self-supervised learning (SSL) methods.

\textbf{SSL in CV} In MOCO \cite{he2020momentum}, SIMCLR\cite{chen2020simple}, BYOL \cite{grill2020bootstrap}, and related works, researchers obtain representations using contrast learning-based methods for self-supervised learning. In BEiT \cite{bao2021beit}, VIT-MAE \cite{he2022masked}, and other works, researchers use MASK patterns for image representation, which is commonly used in NLP. The contrast learning approach relies on a large amount of data augmentation, while the MASK-based approach does the opposite.

\textbf{SSL in NLP} BERT \cite{devlin2018bert} and ALBERT \cite{lan2019albert} use the MLM approach for self-supervised representation learning of natural language, where they mask the input data and use the unmasked data to predict this content. Numerous experiments have shown that these works work well for different downstream tasks. GPT \cite{radford2019language} and other related use autoregressive approaches for representation, which works well in generative tasks. 

\textbf{SSL in MIR}  CLMR \cite{spijkervet2021contrastive} applied multiple novel data augmentation to audio, and used a SimCLR-like \cite{chen2020simple} approach for self-supervised learning. After the great success of the jukebox \cite{dhariwal2020jukebox} in music audio generation, Rodrigo et al. \cite{castellon2021codified} used it for MIR, pre-trained with codified audio language modeling, and for other downstream tasks. S3T \cite{zhao2022s3t} uses the powerful Swin Transformer \cite{liu2021swin} in CV to characterize the spectrum as a pure picture. These works have excellent results but do not combine both the local spatial features of the spectrum and the overall temporal features. Currently, Mulap \cite{manco2022learning} considers these features. However, it introduces additional text information and only embedding of local features. The model proposed by Won et al \cite{Won2021SemisupervisedMT}. is very similar to ours. However, it does not take into account the pre-training task. We aim to extract the features inside the patch without referring to other external data in the pre-training task and to handle the relationship between the sequences of different patch features.

\section{Experiment}
\section{APPROACH}
\label{sec:pagestyle}

This work draws heavily from ViT-MAE \cite{he2022masked}, where the structure of the feature extractor and bottleneck for encoding is more complex than that of the decoder because the main task of learning features is the encoder. Our objective is to enable the encoder part to learn the features of the spectrogram by using the self-supervised learning method of the masked language model \cite{devlin2018bert}. In contrast, the feature encoder and decoder can efficiently learn the interior features inside the patch.

\subsection{Architecture}

The network structure consists of 3 parts: feature extractor, bottleneck, and decoder, where the feature extractor and bottleneck are encoder parts. The feature extractor is for extracting patch features, consisting of a multi-layer convolutional structure similar to VGGnet; the bottleneck is a multi-layer transformer encoder used for sequence feature analysis. Similar to ViT-MAE, our decoder is much lighter than the encoder, consisting of only a few layers of upsampling + conv. 

\textbf{Feature Extractor} Similar to the VGGnet family, a cascaded 3x3 convolution is used instead of a large convolution kernel. In this way, the representational power of the feature extractor can be increased, and the parameters can be reduced moderately. Meanwhile, Springenberg et al. \cite{springenberg2014striving} proposes that converting the first conv layer of the cascade to a conv with stride = 2 can reduce the computational effort and improve the stability. So we also eliminate pooling and use stride conv. Like most of the work, we also use batch normalization to accelerate convergence and alleviate overfitting.

\textbf{Bottleneck} is a relatively lightweight bert encoder. The extended audio sequence is split into several non-overlapping patches $\mathcal{A} = \{a_{1}, a_{2}, ..., a_{n} \}$, $a_{1} \cap a_{2} \cap ... \cap a_{n} = \mathcal{A}$.  For each patch, feature extractor embeds it to a latent vector $v_{i}$, the input to the bottleneck is the features extracted by FE $\{v_{1}, v_{2}, ..., v_{n}  \}$.Since convolutional networks cannot handle masked tokens directly \cite{he2022masked}, we just mask vectors learned from the feature extractor.

\textbf{Decoder} In our model, the decoder is extremely simple compared to the encoder, consisting of only multiple layers of upsampling and convolution. The purpose of the decoder is only to reconstruct the features of the original spectrogram. It is worth noting that, unlike \cite{bao2021beit}, our decoder is also trained in stage II. Because BEiT's decoder learns discrete tokens, while our decoder reconstructs the pixel level of the spectrogram.
\begin{figure*}
	\centering
    \includegraphics[width=1\textwidth]{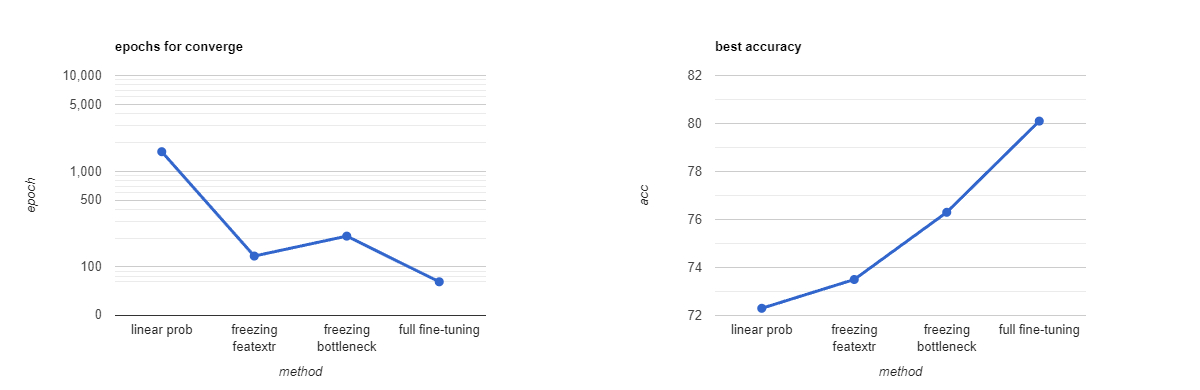}
    \caption{The chart on the left shows the convergence speed of each fine-tuning method, with the learning rate 1e-5. The right is the best accuracy of each. Notice that we use the CLS token for classification tasks. }
    \label{fig:result}
\end{figure*}
\subsection{Two Stage Pre-training}
In order to make the bottleneck learn meaningful features, we adopt a two-stage pre-training pattern. The first stage uses the autoencoding training format to make the feature extractor and decoder initially learn the features inside the patch. The second stage uses MLM to make bottleneck learn the interrelationship between patch sequences. 

\textbf{Stage I} In the first stage, we want the feature extractor and decoder to reconstruct the information of the patch of the random sample, and we DO NOT use the bottleneck part. For each Spectrogram $\mathcal{S}$, we random sample a determined size patch $s, s\in \mathbb{R}^{H \cdot W}$, Then, we send $s$ to the feature extractor and the output vector $v$ directly to the decoder to calculate the MSE loss of the reconstructed patch $\hat{s}$ and the original patch $s$.

\textbf{Stage II} In the second stage, we import the feature extractor and decoder models trained in stage I. The training is acted by the method introduced at the bottleneck. Similar to MAE, we have a learnable mask token specifying the missing patch to bottleneck. The masking strategy is uniform random sampling, which avoids the central bias of mask positions and prevents the model from easily inferring the masked token from neighboring patches. The position encoding is consistent with bert, a learnable pos embedding. The process is to slice the spectrogram into equal patches and send them to the feature extractor to obtain the representation vectors. Then mask these vectors with a high ratio, add the position embedding, and send them to the bottleneck. Finally, the decoder reconstructs these patches and finds their MSE loss.

Our goal is not only to evaluate the performance of the whole pre-trained model in downstream tasks, both classification and regression, but also to evaluate the performance of pre-trained tasks in different stages by freezing different modules as partial fine-tuning. 

\subsection{Dataset}
For pre-training, we used a subset of magnatagatune \cite{law2009evaluation}, with about 16,000 recordings of 29 seconds. This data was tagged by many TagATune gamers, with genres ranging from electronic, rock, classic, etc. For each audio, we sampled it with a sampling rate of 22050 hz, and a mel kernel of 64, so, for our spectrogram, the shape is $[1, 64, W], W \gg 64$. For normalization, we multiply all audio by 0.01 so that all pixels take values in the range [-1, 1]. For downstream, we choose GTZAN \cite{tzanetakis2002musical} as the dataset for genre classification, consisting of 1000 audio tracks of 30 s duration, divided into ten genre classes. For sentiment classification, we used the Emo Music \cite{10.1145/2506364.2506365} dataset, which consists of 744 45-second clips, The authors of JukeMIR \cite{castellon2021codified} have artist-stratified split them, and we have used their code to do the same. The sentiment classification is a regression task, scoring in arousal and valence. Similar to other work, we use the coefficient of determination for evaluation.

\subsection{Setup}
\textbf{Input} For each audio, we sampled 22050. Then we transform to mel spectrogram with filter bank of 64 and hop size of 512. To improve the generalization ability of the model, we did the following data augmentation:
\begin{itemize}
\item Random Scaling: We randomly randomize each spectrogram's size horizontally and the ratio [0.9, 1.2]. Spectrogram, we horizontally randomly resize its size in the ratio [0.9, 1.2]. 
\item Random Crop:  We randomly crop 25 patches per spectrogram with a size of (64, 32). 
\item Random Intensity: For each spectrogram, we randomly element-wise multiply a ratio of [0.8, 1.2] to vary the intensity of each element.
\end{itemize}

\textbf{Model Setup} For the feature extractor, we consist of 5 down convolutional modules. The initial number of filters is 16, increasing layer by layer, using stride conv and batch normalization. The bottleneck consists of 8 layers of bert-like transformers, and each transformer has four attention heads, hidden size is 256. The position embedding is also a bert-like learnable embedding in the embedding layer. Between the decoder and bottleneck, there is a layer of MLP to align the hidden shape of the two modules. The decoder is simple, consisting of only five upsampling layers and convolution. The downstream task consists of one layer of the bert pooler and one layer of the Linear layer.

\begin{table}[]
    \centering
    \begin{tabular}{c | c c}
    \hline
    method & GTZAN & Emo Music \\
    \hline\hline \\
    Patchifier  & 72.3 & 48.6 \\
    CLMR \cite{spijkervet2021contrastive}       & 68.6 &  \\
    JukeMIR \cite{castellon2021codified}    & 79.7 & 69.9\\
    MuLaP  \cite{manco2022learning}    &      & 58.5 \\
    S3T  \cite{zhao2022s3t}      & 81.1 & \\
    \hline
    \end{tabular}
    \caption{Patchifier's classification and regression performance compared to related pre-trained model. Score for GTZAN is percentage of accuracy, while r2 score multiply 100 for Emo Music}
    \label{tab:my_label}
\end{table}
\textbf{Pretraining Setup} We only use a 3070 GPU for pretraining, with a training time of 22 hours and 1000 epochs. Batch size is 32. adamW optimizer is used. LR is 1e-4.

\subsection{Result and Discussion}
We have done partial fine-tuning, comparing the following fine-tuning methods: linear probe, freezing feature extractor, freezing bottleneck, and fully fine-tuning. The differences between the different methods are enormous. As shown in \ref{fig:result}, in the classical gtzan genre classification dataset, the linear probe has the slowest convergence and the worst results. Probably because the data gap between upstream and downstream tasks is relatively large, the pre-training task has various forms of data, and the sampling frequency is not the same as the downstream task. Moreover, linear probe loses the ability to learn more non-linear features \cite{he2022masked}. Freezing feature extractor has the second fastest convergence but the second worst result. The reason may be that the difference in data distribution makes the embedding learned by the feature extractor challenging to generalize in the downstream task. The performance of the Freezing bottleneck is close to full fine-tuning, but the convergence speed is slightly slower. It indicates that our stage 2 pre-training model is significant since the sequence features learned by bottleneck can be well applied to the downstream task.

We compared with other similar work on both gtzan and emo music datasets. We only use the linear probe as a fine-tuning method, and although we use a lightweight model with a limited pre-training dataset size, we still surpass the CLMR baseline. However, there is still a big gap with large-scale models like JukeMIR and S3T. \ref{tab:my_label} shows the results compared with related work. 

Our pre-training process has outstanding reconstruction results even with a large masked ratio. For downstream tasks, our model also performs well. Since we have a tiny amount of data compared to other related work, only 16k audios, we still have a big gap with the state-of-the-art. However, we outperformed CLMR on the gtzan data, which used ten times more data than we did. However, in the regression task, our performance is relatively poor, with a significant gap with related work on emo music. 

\section{CONCLUSION}
In our work, we present Patchifier, a model that treats the audio spectrograms of music as a sequence of large patches and combines methods from CV and NLP for pre-training and several downstream tasks. Besides, we compare different fine-tuning methods and evaluate the effects of different modules and pre-training tasks. We also compare different fine-tuning methods and evaluate the effects of different modules and pre-training tasks. Finally, we provide a novel idea for the future direction of MIR and audio machine learning in general.

\section{Future Work}
We are currently severely short of data and will pre-train the model on large data sets in the future. As shown by partial fine-tuning, the features learned by the feature extractor do not work well for downstream tasks. We will use the full transformer model for pre-training in the future. The current Patchifier only trains on fixed-length data, but the music data is variable-length, and we will pre-train on variable-length data in the future.


\vfill\pagebreak

\bibliographystyle{IEEEtran}
\bibliography{my_refs}

\end{document}